\documentclass[onecolumn,amsfonts,amssymb]{revtex4}
\usepackage{graphicx}
\usepackage{epstopdf}
\usepackage{dcolumn}
\usepackage{amsmath}
\usepackage{epsfig}
\usepackage{indentfirst}
\usepackage{psfrag}
\usepackage{subfigure}
\usepackage{amssymb}
\usepackage{color}

\begin{document}

\renewcommand{\Re}{\operatorname{Re}}
\renewcommand{\Im}{\operatorname{Im}}

\title{Charged-state dynamics in Kelvin probe force microscopy}
\author{Martin Ondr\'{a}\v{c}ek$^{1}$}
\author{Prokop Hapala$^{1}$}
\author{Pavel Jel\'{i}nek$^{1}$}
\affiliation{
$^1$Institute of Physics  of the Czech Academy of Sciences, Cukrovarnick\'a 10/112, 162 00 Prague, Czech Republic
}

\begin{abstract}
We present a~numerical model which allows us to study the Kelvin force probe microscopy response to the charge switching in quantum dots at various time scales. The model provides more insight into the behavior of frequency shift and dissipated energy under different scanning conditions measuring a~temporarily charged quantum dot on surface. Namely, we analyze the dependence of the frequency shift, its fluctuation and of the dissipated energy, on the resonance frequency of tip and electron tunneling rates between tip--quantum dot and quantum dot--sample.
We discuss two complementary approaches to simulating the charge dynamics, a stochastic and a deterministic one. In addition, we derive analytic formulas valid for small amplitudes, describing relations between the frequency shift, dissipated energy, and the characteristic rates driving the charging and discharging processes. 
\end{abstract}

\date{\today}
\maketitle

\section {Introduction}
Further development of electronic devices and their performance strongly depends on our ability to characterize and control charge distribution down to atomic scale. From this perspective, scanning tunneling microscopy (STM) \cite{Binnig_PRL1982} and atomic force microscopy (AFM) \cite{Binnig_PRL1986} are very important tools that allow imaging and manipulation of single atoms on surfaces.  Namely, recent development of dynamical atomic force microscopy (dAFM) \cite{Albrecht_JAP1991, ncafm_vol3,Giessibl_RMP2003} provides a powerful tool to probe the local structure \cite{Gross_Science2009, deOteyza_Science2013,Emmrich_Science15, Hapala_PRB2014} and chemical composition \cite{Sugimoto_Nature07,Setvin_ACSNano12} of surfaces  with a resolution reaching the atomic scale on semiconductor \cite{Giessibl_Science95}, metallic \cite{Loppacher_PRB00} and also insulating \cite{Barth_Nature01} surfaces.
Kelvin Probe Force Microscopy (KPFM), a~technique  derived from dAFM,  can be used to probe local variations of local work functions down to nanometer scale \cite{Nonnenmacher_APL91}. It has been demonstrated that KPFM is able to reach the atomic  \cite{Sadewasser_PRL09, Nony_PRL09,Enevoldsen_PRL08}and sub molecular \cite{Mohn_NatNano2012} resolution too. The straightforward interpretation in terms of local contact potential difference is no more possible at the atomic scale though \cite{Schuler_NanoLett2014,Albrecht_PRL2015,Neff_PRB15,Corso_PRL15}.

Recent development of scanning probe technique (see e.g.~\cite{Loppacher_PRB00,Giessibl_APL98,Sugimoto_PRB10}) allowed simultaneous acquisition of AFM and STM channels. This brought new possibilities not only for advanced characterization of  surfaces \cite{Majzik_ACSNano2013,Baumann_ACSNano14} but also to study an~influence of the tunneling current on detected forces \cite{Weymouth_PRL11}.
The latter relies on a~proper decoupling of AFM and STM channels \cite{Majzik_BJN11}.  Kelvin probe force microscopy and spectroscopy (KPFM/KPFS) complemented with STM channel have the potential to become a convenient tool for studying the dynamics of charging and discharging local structures here referred as ``quantum dots'' (QD). These QDs are thought of as objects capable of storing electric charge and thus being able to switch between two or more distinguishable charge states (for instance charged and neutral).
They may have the form of nanoclusters (like InAs clusters on InP in Ref. \cite{Stomp_PRL05, Cockins_PNAS10, Bennett_PRL10}), molecules or even single atoms (like Au atoms on NaCl in \cite{Repp_Science04,Gross_Science2009a}) 

Such QDs are of considerable interest because of their potential applications in nanoelectronics \cite{Ray_NatNano08}.
Above referenced examples have been indeed successfully probed using KPFM in two seminal papers \cite{Stomp_PRL05,Gross_Science2009a}, which paved a route towards a new concept of controlling charge on atomic scale. However, the characteristic time scales of this quantum-dot charge-state dynamics can span many orders of magnitude depending on the details of the studied system as well as on the immediate position of the KPFM scanning probe.
Consequently, the time scales of the charge dynamics relate in different ways to the characteristic time scales of the measuring instrument,
like the cantilever oscillation period $T_o$, reaction time of the electronic feedback of the AFM machine, acquisition time for a single point in the measurement etc.
The manifestation of phenomena related to the switching in the KPFM data will depend on the ratio of the various time constants involved. This fact may complicate interpretation of the experimental results.

The aim of this paper is to develop a~numerical model which allows us to simulate the KPFM response to the charge switching in QD at various time scales. 
While gating effect 
of the biased AFM tip was crucial for the explanation of charging in the above quoted experiments on Au atoms or InAs nanostructures \cite{Repp_Science04,Gross_Science2009a,Stomp_PRL05,Kocic},
our focus in this paper will be on cases where an~electron tunneling from the tip has the decisive effect. We should note that the model could be easily adapted to the charge gating processes too, but this is beyond the scope of the paper. We develop and compare two complementary approaches, \emph{stochastic} and \emph{deterministic}, to simulate the charge dynamics. 
Such insight into the response of oscillating probe driven by the charging processes helps in better understanding the complex behavior observed in KPFM experiments. In addition, we will establish a relatively simple analytic formulas providing, in the limit of small amplitudes, relations between frequency shift $\Delta f$, dissipated energy $E_{\mathrm{diss}}$, and characteristic tunneling rates ($\nu_1$, $\nu_2$) driving the charging processes.

\section{Model}

\begin{figure}
\includegraphics[width=\textwidth]{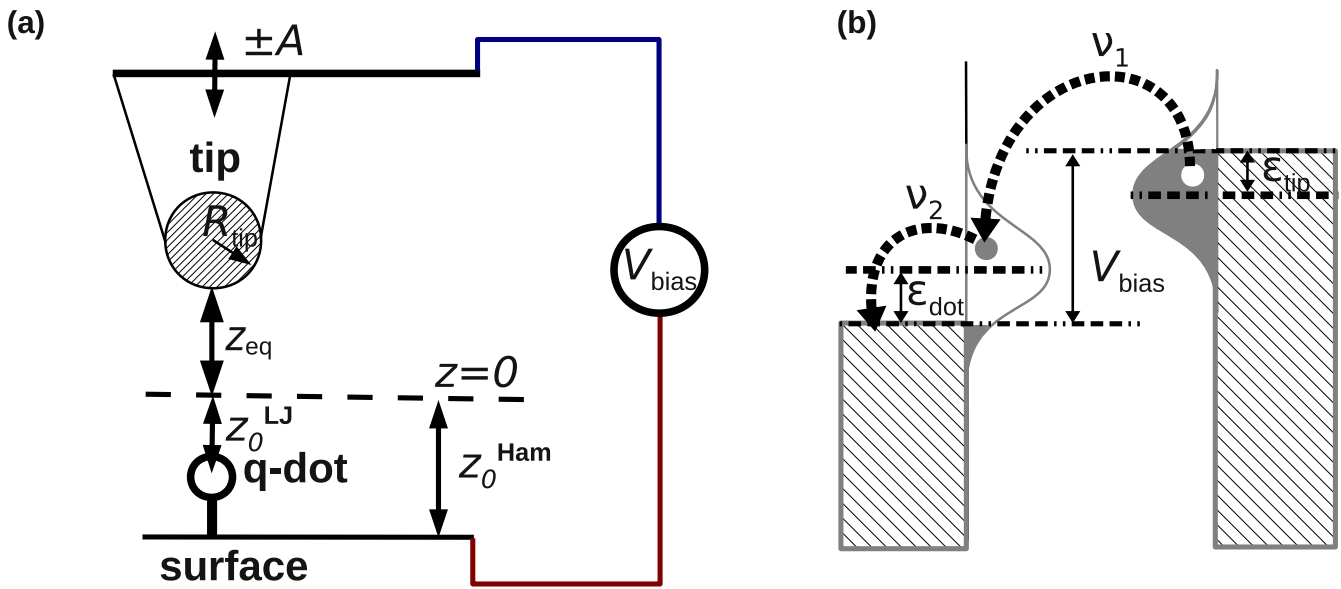}
\caption{\label{Fig_scheme}
(a) Schematic sketch of the model used in this paper. The model consists essentially of a tip on a cantilever and a QD on a surface.
The geometry of the model is in fact treated as purely one-dimensional, i.e.\ only the $z$ coordinate is relevant.
(b) Model of the tunneling process between the tip and the QD. The tunneling from the tip to the dot is assumed to be elastic and two localized states are assumed to take part in it,
one at the tip termination and the other one in the dot, resulting in a Gaussian-shaped resonance.
}
\end{figure}

The outline of the model we propose (sketched in Fig~\ref{Fig_scheme}(a)) is as follows.
Consider a QD (be it an atom, molecule, or a nanostructure) on a surface.
Assume the dot can accept or release an electron, thus switching between two possible charge states.
The AFM probe is sensitive to the charge state of the QD due to the electrostatic component of force between the charged dot and the probe apex.
Several time 
constants or rate parameters determine the characteristic time scales of the model:
\begin{itemize}
\item Tunneling rate between the AFM tip and the QD on the surface $\nu_1$.
\item Discharging rate of the charged dot by tunneling to a conductive substrate $\nu_2$.
\item Oscillation period of the AFM tip $T_o = 1/f_0$, where $f_0$ means resonant oscillation frequency .
\item Measuring time $t_m$.
\end{itemize}
Furthermore, the tunneling probability between the tip and the dot depends on tip position ($z$) and on applied voltage bias ($V$).

\subsection{Forces}
We aim to create a model as simple as possible but still capturing the essential features of charge dynamics coupled to tip oscillations.
The force between the AFM tip and the surface consists of the following components:
\begin{itemize}
\item $F_{\mathrm{LJ}}$ derived from the Lenard-Jones potential between the QD and the probe apex
(represents the short-range forces independent from charge and bias: Pauli repulsion and the local part of van der Waals interaction).
\begin{equation}\label{LJforce}
F_{\mathrm{LJ}} (z) =  12 E_{\mathrm{min}}^{\mathrm{LJ}} \left( \frac{(R_{\mathrm{min}}^{\mathrm{LJ}})^{12}}{\left( z - z_0^{\mathrm{LJ}} \right)^{13}} - \frac{(R_{\mathrm{min}}^{\mathrm{LJ}})^{6}}{\left( z - z_0^{\mathrm{LJ}} \right)^{7}} \right),
\end{equation}
\item $F_{\mathrm{Ham}}$ derived from the Hamaker model for a spherical tip of radius $R_{\mathrm{tip}}$ and planar surface \cite{Argento} to describe the long-range part of van der Waals (or, more precisely, London dispersion) interaction.
The Hamaker model approximates the tip with a homogeneous sphere and the surface and substrate beneath it by a homogeneous half-space delimited by the surface plane.
\begin{equation}\label{HamakerForce}
F_{\mathrm{Ham}}(z) = -\frac{A_H R_{\mathrm{tip}}}{6(z - z_0^{\mathrm{Ham}})^2}.
\end{equation}
\item The electrostatic part $F_{\mathrm{cap}}+F_Q$. We describe the long-range part of the electrostatic force $F_{\mathrm{cap}}$ as the force between two electrodes of a capacitor,
one of which is planar and represents the surface and other one is spherical and represents the tip \cite{Hudlet}.
The same geometry, that is the same tip radius and distance from the surface, is assumed as in the Hamaker model used above.
To calculate the local part of the electrostatic force $F_Q$, we treat the charge of the QD as a point charge of the size corresponding to one electron.
The point charge $q$ may be present or absent, depending on the immediate ``charge state''.
If absent, the local electrostatic contribution is assumed to be zero. If present, we assume an interaction of the point charge with
(i) an electric field from the tip and (ii) an other static point charge $Q_{\mathrm{apex}}$ positioned on the very end of the AFM tip.
The electric field is assumed to depend linearly on the voltage bias and be inversely proportional to the tip distance from the QD (as if the field were homogeneous) while the apex charge is assumed to be characteristic for the tip and independent from the bias.
\begin{equation}\label{CapacitiveForce}
F_{\mathrm{cap}} (z,V) = -\frac{\pi \epsilon_0 R_{\mathrm{tip}}^2 (V - V_{\mathrm{CPD}})^2}{(z - z_0^{\mathrm{Ham}}) (R_{\mathrm{tip}} + z - z_0^{\mathrm{Ham}})}
\approx -\frac{\pi \epsilon_0 R_{\mathrm{tip}} (V - V_{\mathrm{CPD}})^2}{z - z_0^{\mathrm{Ham}}},
\end{equation}
where $\epsilon_0 = 8.854187817\times 10^{-12}$~F/m is the vacuum permittivity,
and
\begin{equation}\label{CoulombForce}
F_Q (z,V) = \frac{e_0^2}{4\pi \epsilon_0} \frac{q Q_{\mathrm{apex}}}{(z - z_0^{\mathrm{LJ}})^2} + \frac{e_0}{2} \frac{q (V - V_{\mathrm{CPD}})}{z - z_0^{\mathrm{LJ}}},
\end{equation}
where $e_0 = 1.6021765\times 10^{-19}$~C is the elementary charge. The charges $q$ and $Q_{\mathrm{apex}}$ are expressed here in the unit of the electron charge (negative elementary charge), so they are themselves dimensionless quantities.
\end{itemize}
The total force by which the surface with the QD pushes the tip is then $F(z,V)=F_{\mathrm{LJ}}(z) + F_{\mathrm{Ham}}(z) + F_{\mathrm{cap}}(z,V) + F_Q(z,V)$.

The mechanical dynamics of the tip is characterized by its (undamped) resonant oscillation frequency $f_0$, cantilever stiffness $k$, and quality factor $Q$.
For the instantaneous vertical position $z(t)$ of the tip, the equation of motion
\begin{equation}
\frac{d^2 z}{dt^2} + \frac{2\pi f_0}{Q} \frac{dz}{dt} + (2\pi f_0)^2 (z - z_{\mathrm{eq}}) = F(z)
\end{equation}
holds, where $z_{\mathrm{eq}}$ is the equilibrium position of the tip determined by the piezolectric positioning system of the AFM cantilever
and $F$ is the total force applied on the oscillating tip by its interaction with the surface below it. The quality factor $Q$ and resonance frequency (in the absence of damping) $f_0$ characterize the cantilever.

\subsection{Model for the electronic states of the tip apex and on the surface}
The electronic state at the QD, to which or from which the electrons are tunneling when the charge state changes,
is characterized by its mean energy   $\epsilon_{\mathrm{dot}}$ and a broadening $ w_{\mathrm{dot}}$ around this mean energy $\epsilon_{\mathrm{dot}}$.
The broadening $ w_{\mathrm{dot}}$ is a consequence of an interaction with the substrate.
We take the shape of the energy broadening is described by Gaussian distribution (but the exact shape is not crucial for the effects described in this paper) then the density of state of QD $\rho_{\mathrm{dot}}$ holds:

\begin{equation}\label{gaussdot}
\rho_{\mathrm{dot}} = \frac{1}{\sqrt{\pi} w_{\mathrm{dot}}} \exp \left ( \frac{\epsilon - \epsilon_{\mathrm{dot}}}{w_{\mathrm{dot}}} \right)^2.
\end{equation}

Two models of the electronic structure of the tip were implemented: i) either the tip carries a single electronic state capable of participation in the tunneling (this model may represent a tip with a localized state, like a dangling bond, on its terminal apex);  or ii) the tip hosts a continuum of electronic states with a constant density of states near the Fermi level (a metallic tip). 
In this paper, we discuss only the localized-state tip model as the continuum model gives very similar results (see discussion later).
In the case of the tip model with the localized apex state, we describe the corresponding density of states on the tip $\rho_{\mathrm{tip}}$ by a Gaussian distribution, similarly to the QD state:

\begin{equation}\label{gausstip}
\rho_{\mathrm{tip}} = \frac{1}{\sqrt{\pi} w_{\mathrm{tip}}} \exp \left ( \frac{\epsilon - \epsilon_{\mathrm{tip}}}{w_{\mathrm{tip}}} \right)^2.
\end{equation}

Fermi-Dirac distribution of occupancy probabilities is invoked to account for a finite temperature of the tip $T$. Temperature effects on the substrate are disregarded. The tunneling probability per unit time from the tip to the QD ($\nu_1$) is assumed to exponentially decay with tip distance from the surface.
Further, this tunneling probability $\nu_1$ resonantly depends on voltage bias through the Gaussian densities of states on the dot and tip in a way that corresponds to the assumption of exclusively elastic tunneling. Taking all these conditions in account, the tunneling probability from the tip to QD is expressed as:
\begin{equation}\label{tunneling}
\nu_1 (z, V) = \nu_0 \exp(-\beta z) N_{\mathrm{norm}} \int_{0}^{\infty} d\epsilon \rho_{\mathrm{dot}} (\epsilon) \rho_{\mathrm{tip}} (\epsilon - e_0V) f_{\mathrm{FD}} (\epsilon - e_0V, T).
\end{equation}

Fig.~\ref{Fig_scheme}(b) displays a scheme of the tunneling process.
In the model presented here, we do not consider the possibility of tunneling in the opposite direction, i.e.\ from the dot to the tip. While this may seem to be an unrealistic assumption,
as the probability of elastic tunneling should be the same in both direction, it may in fact be a reasonable approximation for cases similar to Au on NaCl.
The negative Au$^{-}$ ion, once created, gets stabilized by interaction with the substrate and does not change back to its original neutral state at the same tip bias \cite{Repp_Science04}.
In our model, the charged dot can resume its neutral state by giving up an electron to the substrate. The probability of this event in a unit of time is a constant parameter ($\nu_2$) of the model.
The normalization factor $N_{\mathrm{norm}}$ in Eq.~(\ref{tunneling}) is chosen in such a way that the tunneling rate at $V = (\epsilon_2 - \epsilon_1)/e_0$ and $z=0$ is $\nu_1 = \nu_0$.
The prefactor $\nu_0$ is a $z$- and $V$-independent parameter, which characterizes the overall feasibility of tunneling from the tip to the dot.

\subsection{Stochastic vs deterministic model}

We compare two approaches to simulating the charge dynamics, to which we will refer as a \emph{stochastic} and a \emph{deterministic} approach.
In both cases the probabilities of electron tunneling from the tip to the dot ($\nu_1 dt$) and from the dot to the substrate ($\nu_2 dt$)
are evaluated in each time step for the present position of the tip. In the stochastic case, they are treated as genuine probabilities and a pseudorandom number generator is used to decide
whether the charge state of the QD should be changed or not based on these probabilities and the pseudorandom value.
Only charges 0 or 1 (meaning no or one extra electron) were allowed in the stochastic approach. When 0, it changes to 1 in the next step with probability $\nu_1 dt$; when 1, it changes to 0 with probability of $\nu_2 dt$.
In the deterministic approach, the tunneling rates $\nu_1$ and $\nu_2$ are interpreted as a charge flow. A fractional charge between 0 and 1 is allowed and the current value of charge $q(t)$
is updated in every time step according to the rate equation
\begin{equation}
q(t + dt) = q(t) + \nu_1 (z(t), V) (1- q(t)) dt - \nu_2 q(t) dt.
\end{equation}
This finite-step evolution of $q(t)$ corresponds, in the $dt \rightarrow 0$ limit, to the differential equation (in simplified notation omitting the functional dependences)
\begin{equation}\label{Qrate}
\frac{dq}{dt} = \nu_1 (1-q) - \nu_2 q.
\end{equation}

All simulations of the tip dynamics were carried out using following parameters: $z_{\mathrm{eq}} = 4.0$~\AA{}, $f_0 = 46.858$~kHz, $k = 3681$~N/m, $Q = 2438$ and a small amplitude of $A = 0.1$~\AA{}. The curves were sampled with the step of $\Delta V = 0.02$~V. The bias voltage $V$ was kept constant for 1000 oscillation periods of the cantilever to measure one point on a $\Delta f (V)$ curve.
Following values were chosen for the parameters that determine individual force components in our model:
$E_{\mathrm{min}}^{\mathrm{LJ}}=0.2$~eV, $R_{\mathrm{min}}^{\mathrm{LJ}}=3$~\AA{}, $z_0^{\mathrm{LJ}}=-3.2$~\AA{}, $R_{\mathrm{tip}}=400$~nm,  $A_H =0.029$~eV, $z_0^{\mathrm{Ham}}=-4.8$~\AA{}, $V_{\mathrm{CPD}}=0$,
$Q_{\mathrm{apex}}=-0.5$.
The decay constant for tunneling probability was $\beta$=2.3~\AA$^{-1}$.
This choice of parameters is mostly arbitrary but should be quite realistic, as it was motivated by several (unpublished) STM/AFM measurements carried out in our department's lab by our experimentalist colleagues.

\section{Results}
\begin{figure}
\includegraphics[width=0.85\textwidth]{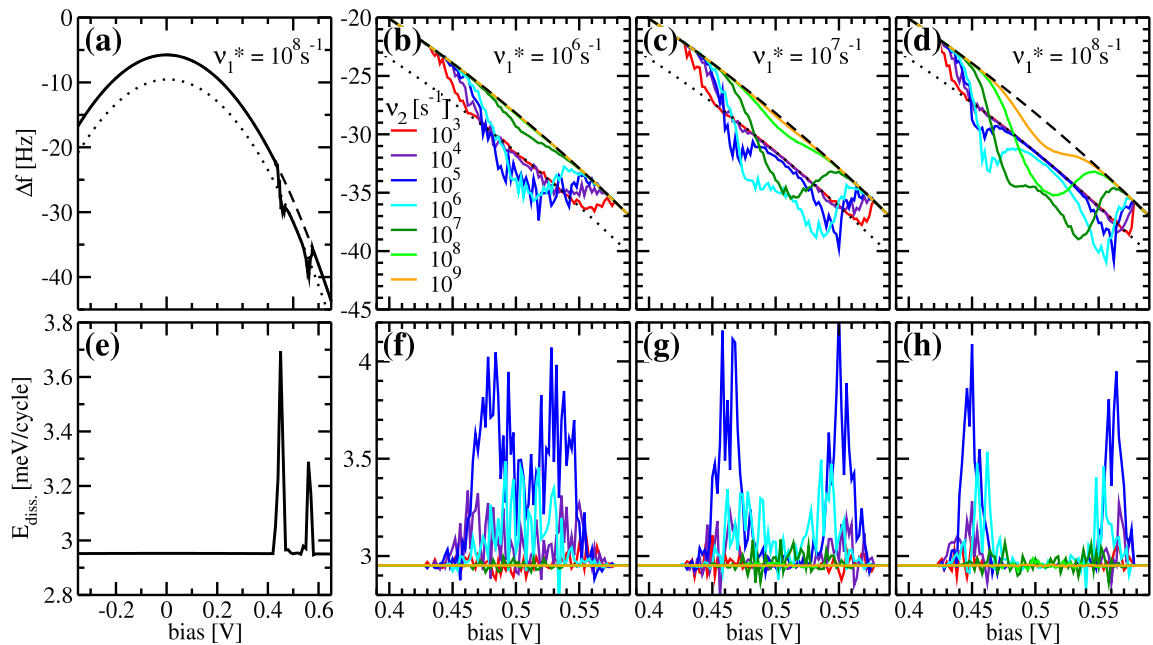}
\caption{\label{dfResults}
(a,e) $\Delta f$ and $E_{\mathrm{diss}}$ for $\nu_1^* = \nu_0 \exp(-\beta z_{\mathrm{eq}}) = 10^8$~s$^{-1}$, $\nu_2=10^5$~s$^{-1}$ and $f_0 = 4.6858\times 10^4$~Hz in a wider range of voltage to illustrate the typical shape of whole measured curves including the parts away from the resonance.
(b-d) A series of simulated $\Delta f (V)$ measurements for different tunneling rate prefactors $\nu_0$ (specific for each panel, expressed as $\nu_1^*$), and different QD discharging rates $\nu_2$ (encoded by the color of the corresponding curve), with the same $f_0$.
(f-h) Corresponding series of simulated results for the dissipated energy per oscillation (or, equivalently, power needed to drive the oscillation) as a function of voltage bias.
(b,f) $\nu_1^*=10^6$~s$^{-1}$, (c,g) $\nu_1^*=10^7$~s$^{-1}$, (d,h) $\nu_1^*=10^8$~s$^{-1}$,
$\nu_2$ ranges from $10^3$~s$^{-1}$ (red), through $10^4$~s$^{-1}$ (violet), $10^5$~s$^{-1}$ (dark blue), $10^6$~s$^{-1}$ (light blue), $10^7$~s$^{-1}$ (dark green), $10^8$~s$^{-1}$ (light green) to $10^9$~s$^{-1}$ (orange).
}
\end{figure}

Fig.~\ref{dfResults} presents results of the stochastic model for various charging and discharging rates determined by parameters $\nu_0$ and $\nu_2$. The values of $\nu_0$ are indicated indirectly, in terms of $\nu_1^* = \nu_0\ \exp(-\beta z_{\mathrm{eq}}) \approx 10^{-4} \nu_0$.
Such rescaling by the exponential factor facilitates a direct comparison between the different time scales.

The figure displays a set of curves that represent the simulated frequency shift $\Delta f$ and the dissipated energy $E_{\mathrm{diss}}$ as a function of voltage bias $V$.  Occurrence of the resonant tunneling though a~localized QD electronic state $\epsilon_{\mathrm{dot}}$  introduces a~variation of the frequency shift $\Delta f$ at the corresponding voltage, as seen in Fig.~\ref{dfResults}(a). This effect is accompanied with an~appearance of a pronounced signal in the dissipated energy $E_{\mathrm{diss}}$ channel.
Here we should note that similar phenomena at bias voltages near the threshold for tunneling are also observed with the metallic-tip model, which is not discussed here any further.
The only difference between results obtained with a localized-state and a metallic tip is that in the latter case the system stays in the charged state for higher voltages as a consequence of the continuum density of tip states $\rho_{\mathrm{tip}}$.

Both the variation of the frequency shift $\Delta f$ and the dissipated energy $E_{\mathrm{diss}}$  is directly related to the charging process occurring during the resonant tunneling. 
In our model, the local electronic states in the QD and on the tip apex are chosen so that a resonance for tunneling occurs around the bias of +0.5~V.
The relevant parameters were: the dot state energy $E_\mathrm{dot}=$0.5~V with respect to the substrate  Fermi level, the apex state energy $E_\mathrm{tip}=$0 (right a the Fermi level of the tip),
and Gaussian smearing for both states was given by the full width at half maximum of 0.03~eV (corresponding to $w_{\mathrm{dot}} = w_{\mathrm{tip}} \approx 0.05$~eV).
As expected, the QD stays discharged (neutral) most of the time for out-of-resonance values of voltage bias. In the neutral state, the $\Delta f (V)$ dependence follows a parabolic curve as known from traditional KPFM measurements, see Fig.~\ref{dfResults}(a) dashed line. Note that the dissipation signal is negligible  Fig.~\ref{dfResults}(e).

The situation changes when the tunneling junction is brought to the resonance ($V$ around 0.5~V) and the tip is sufficiently close. Provided the prefactor $\nu_0$ of Eq~(\ref{tunneling}) is sufficiently large too, the tunneling rate between the tip and the QD becomes much larger than the discharging rate, $\nu_1 \gg \nu_2$.
Consequently, the QD will be found 
in the charged state most of the time. In such a case,
the parabolic curve just shifts downwards (or upwards) by a constant amount of $\Delta f$ with respect to the neutral case, depending on the sign of an extra Coulomb force $F_Q$. Here we consider the attractive interaction $F_Q$ so the frequency shift $\Delta f$ is more negative. 

One can expect that the system oscillates between two (neutral and charged) states represented by idealized Kelvin parabolas, one for the charged (dot line) state and the other for the neutral one (dashed line). However, the simulations reveal that the frequency shift does not have to lie always between the two parabolas, see Fig.~\ref{dfResults}(b-d). In other words, the changes of the frequency shift with respect to the neutral state sometimes tend to ``overshoot'' the curve that would correspond to the $\Delta f(V)$ on a fully charged state. We observe this effect, in particular, when the charging and discharging probabilities $\nu_1, \nu_2$ are of comparable magnitude and the QD therefore often changes its charge state.
Such a situation occurs in two different regimes. First, for high resonant tunneling rates $\nu_0$, at the ``edges'' of the resonance, where the high resonant tunneling rate $\nu_0$ becomes partially compensated by being slightly off-resonance. 
Thus it makes the actual tunneling rate $\nu_1$ from the tip comparable to the discharging rate  $\nu_2$.  Second, it happens for moderate resonant tunneling rates  $\nu_0$ just at the center of the resonance. The regions of the overshoot partially overlap with regions of large dissipation, cf.~Fig.~\ref{dfResults}(b-d) and Fig.~\ref{dfResults}(f-h). Nevertheless the dissipation tends to increase for slower charging and discharging rates (closer to the tip oscillation frequency $f_0$) while for fast charging and discharging rates, the dissipation is small even at voltages corresponding to the $\Delta f$ overshoot.

\begin{figure}
\includegraphics[width=0.5\textwidth]{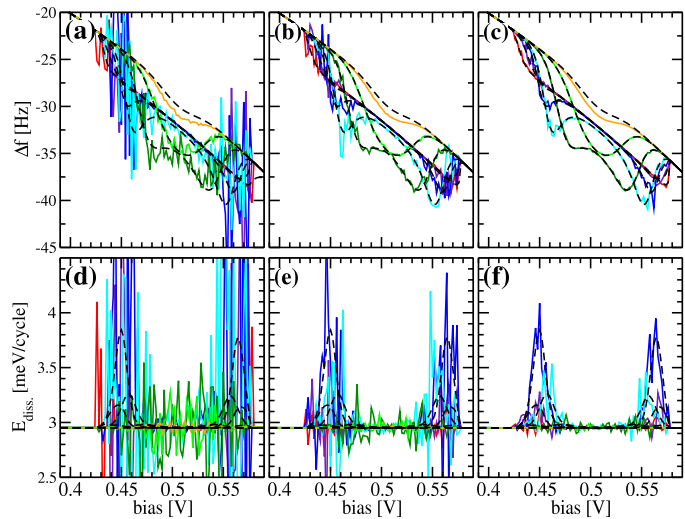}
\caption{\label{StatisticsFig}
Comparison of results between the stochastic and the deterministic approach.
Simulated frequency shift $\Delta f(V)$ (a-c) and dissipated energy $E_{\mathrm{diss}}$ (d-f) are shown for different numbers of oscillation periods over which the result is averaged.
The number of periods taken into the averaging is the total number of periods spend on measuring, viz.\ (a,d) 10 periods, (b,e) 100 periods and (c,f) 1000 periods.
The color code for curves showing the stochastic simulation is the same as in Fig.~\ref{dfResults}, the curves corresponding to the deterministic simulation are all depicted as dashed black lines.
Comparison to the stochastic results can be used to identify curves corresponding to individual $\nu_2$ values in the deterministic case.
}
\end{figure}

Now, we compare the stochastic {\it vs} deterministic approach, shown in Fig.~\ref{StatisticsFig}, to get more insight into the effect of randomness in the charging on AFM dynamics. 
The outcome of the stochastic simulation crucially depends on the number of oscillations spend at each point of the measurement.  In other words, it depends on the time which it takes to measure one point of the $\Delta f(V)$ or $E_{\mathrm{diss}}$ plot.
The longer this measuring time, the better the stochastic results average to suppress fluctuations and approach the mean value expected for the given point.
Fig.~\ref{StatisticsFig} demonstrates that the stochastic simulation agrees perfectly with the deterministic one if the measuring time is long enough, with the exception of the orange curves representing $\nu_2 = 10^9$~s$^{-1}$.
In this case, the discharging to substrate is so highly probable that the mean life time of the charged state is comparable to the simulation step $dt = 0.0001 / f_0 \approx 2\times10^{-9}$~s$^{-1}$, which distorts the results of the simulations.
Even for the shortest measuring time shown, $t_m=10/f_0$ (10 periods), the frequency shift and dissipation calculated in the stochastic way tend to fluctuate around the values given by the deterministic simulation.  However, the fluctuations are large in that case, so they almost mask the functional dependence.
The region of large dissipation at bias near but slightly off the resonance is also a region of largest fluctuations both of the frequency shift  $\Delta f$  and of the dissipated energy $E_{\mathrm{diss}}$.

\section{Theoretical analysis}

First, we will analyze the effect of the (dis)charging process of the QD under the tip  on the measured frequency shift $\Delta f$.   In this paragraph, we provide an~intuitive qualitative explanation of relation between  frequency shift $\Delta f(V)$ and the (de)charging rates $\nu_1, \nu_2$. A~detailed derivation of quantitative formulas for the frequency shift in a small-amplitude limit can be found in Appendix A.  If the system stood in the neutral state indefinitely, the frequency shift $\Delta f(V)$ would follow one Kelvin parabola as sweeping the voltage bias through a finite range; see dashed lines in Fig.~\ref{dfResults}(b-e). Similarly for the charged state, the system follows  a different parabola---see dotted lines in Fig.~\ref{dfResults}(b-e)---which is rigidly shifted according to the neutral case. 
Now, for a suitable voltage, a resonance between the electronic state on the tip apex and the electronic state of the QD is established. This initiates the electron transfer from the oscillating tip into the dot. 

If the rate of QD charging $\nu_1$  is comparable to the rate of discharging $\nu_2$, the average charge on the QD will be something between 0 and 1 electron. This mean value of the charge corresponds accordingly to a frequency shift $\Delta f$ somewhere in between the two above mentioned parabolas. However, the mean charge does not yet explain why the frequency shift sometimes falls outside the area between the two parabolas. In particular, it goes below the lower parabola that corresponds to the fully charged state, as seen in Fig.~\ref{dfResults}(b-e). 

To understand this extra frequency shift, correlation between the dynamics of the charge and tip oscillations has to be taken into account. As the tunneling is more probable when the tip comes closer to the QD during its oscillation, the charge of the QD will be on average closer to 1 when the tip goes through its lower positions. Conversely, the charge will on average be closer to 0 when the tip is in its upper positions. This variation of charge during the tip oscillation creates an~extra effective gradient of the electrostatic force between the tip and the sample, on top of the usual distance dependence arising e.g.\ from the $1/r^2$ factor in the Coulomb law.  Namely, the frequency shift $\Delta f $ derived within the approximation of small oscillation amplitude are given by expression (see also Eqs.~(\ref{DeltaF_final}) in Apendix A):
\begin{equation}\label{DeltaF}
\Delta f(z_{\mathrm{eq}},V) = \frac{f_0}{2k} \left( -\frac{\partial F \bigl(z_{\mathrm{eq}},\bar{q}(z_{\mathrm{eq}},V)\bigr) }{\partial z_{\mathrm{eq}}} + 
\frac{\partial F \bigl(z_{\mathrm{eq}},\bar{q}(z_{\mathrm{eq}},V)\bigr) }{\partial \bar{q}} \frac{\beta \nu_2 \nu_1(z_{\mathrm{eq}},V)}{\bigl( \nu_2 + \nu_1(z_{\mathrm{eq}},V) \bigr)^2 + (2\pi f_0)^2} \right),
\end{equation}
where $A$ is the amplitude of the oscillation, $\beta$ the decay factor of the tunneling probability with distance, $f_0$ a resonance frequency of the cantilever, 
$\nu_1(z_{\mathrm{eq}},V)$ the mean tunneling rate and $\nu_2$ the (constant) discharging rate to the substrate.

The first term on the right-hand side of Eq.~(\ref{DeltaF}) is a contribution to the frequency shift which includes all charge-independent forces as well as the charge-dependent force component evaluated for the time-averaged value of the charge $\bar{q}$. The second term corresponds to the modification of the frequency shift by the charge dynamics.
We can see that the relative frequency shift $\Delta f/f_0$ corresponding to this second term tends to be maximal when $\nu_1 (z_{\mathrm{eq}},V) \approx \nu_2 \gg f_0$.
In such instances, peaks on the $\Delta f(V)$ curve can be expected to appear.

\begin{figure}
\includegraphics[width=0.5\textwidth]{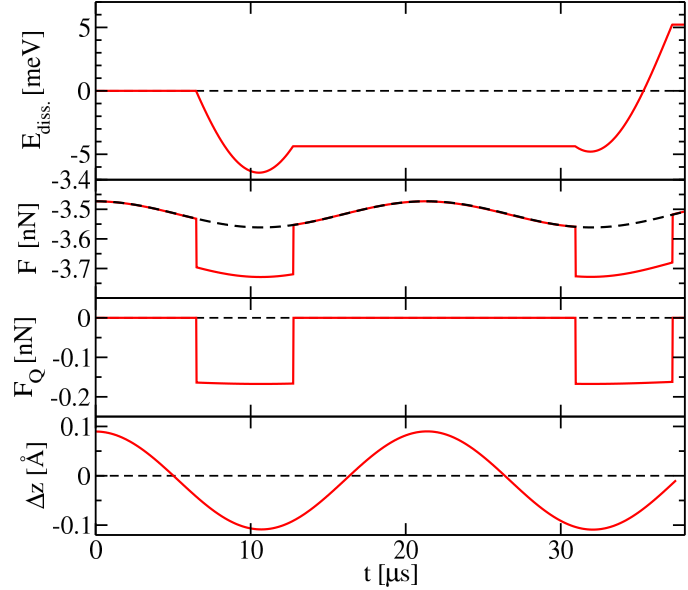}
\caption{\label{Fourier}
Illustration of the relation between charge switching 
and energy dissipation. Panels from bottom to top:
Vertical position of the tip $\Delta z$ as a function of time $t$; charge-dependent force component $F_Q(t)$; total force $F(t)$ acting between the tip and the sample (dashed line shows the charge-independent part); and the energy $E_{\mathrm{diss}}$ dissipated since $t=0$
($dE_{\mathrm{diss}} = F dz$).
}
\end{figure}

Similarly, correlation between the temporal charging and probe dynamics is also manifested by appearance of enhanced signal in the  energy dissipation channel\cite{Stomp_PRL05,Cockins_PNAS10} $E_{\mathrm{diss}}$. 
It means that energy has to be supplied to the cantilever (or sometimes retrieved from it, if the dissipation is negative) in order to maintain constant amplitude $A$ of the oscillation.
From the results shown in Fig.~\ref{StatisticsFig} we can see that the dissipation signal appears at the ``edges'' of the resonance.

Because of the stochastic nature of the charge dynamics, random fluctuations in the measured values of both the frequency shift and energy dissipation have to be expected,
as exemplified by the results of the stochastic simulations plotted in Fig.~\ref{StatisticsFig}. Such fluctuations are particularly pronounced when each measurement involves only few oscillation cycles.
Quantitative expressions for this fluctuations will be derived for the small-amplitude approximation in Appendix C.
Interesting observation can be made if we express the magnitude of frequency-shift fluctuations as the relative root mean square deviation from the mean expectation value of $\Delta f$.
This quantity turns out to be proportional (as a function of the three time scales $f_0$, $\nu_0$ and $\nu_2$) to the square root of the expectation value of the energy dissipation $E_{\mathrm{diss}}$:
\begin{equation}\label{FluctAndDissip}
\delta \left( \frac{\Delta f}{f} \right) = \frac{\sqrt{\partial F/\partial q}}{\sqrt{2N}\pi kA^2 \sqrt{\beta}} \sqrt{\left|E_{\mathrm{diss}}\right|}.
\end{equation}
See Appendix~C for the detailed derivation of Eq.~(\ref{FluctAndDissip}) and for the definition of all quantities that appear in it.
Eq.~(\ref{FluctAndDissip}) suggests that one should expect large fluctuations in the measured value of $\Delta f$ whenever the dissipated energy $E_{\mathrm{diss}}$ is also large.
Indeed, such effect is observed in our numerical simulations, see Fig.~\ref{StatisticsFig}. The instabilities of the frequency can be understood as consequence of the frequent abrupt changes of the force $F_Q$ which the tip experiences during its oscillation.
Based on on Eq.~(\ref{FluctAndDissip}) together with the analysis of $E_{\mathrm{diss}}$ that will follow, we realize that large frequency-shift fluctuations are expected to appear under the condition  $\nu_1 \approx \nu_2 \approx f_0$.
Moreover, the presence of large fluctuation in the frequency shift channel is accompanied by large fluctuations of the dissipated energy $E_{\mathrm{diss}}$ too.

To get more insight into the origin and character of the dissipation signal $E_{\mathrm{diss}}$ during the (de)charging process, let us analyze the dynamics of the probe driven by time-dependent Coulomb force $F_Q$.
Fig.~\ref{Fourier} illustrates correlation between the dissipation signal $E_{\mathrm{diss}}$ and time-dependent force $F_Q$ during two tip oscillation periods. 
The cosine function in the bottom panel represents the immediate position $\Delta z(t) = z(t) - z_{\mathrm{eq}}$ of the tip with respect to its equilibrium  position $z_{\mathrm{eq}}$.
The second panel from the bottom shows what the time development of the Coulomb force $F_Q(t)$ may look like. The force $F_Q(t)$ directly relates to the charge $q(t)$ of the QD.
By definition, the Coulomb force equals zero ($F_Q = 0$) for $q=0$ and it jumps to a non-zero value when the state of the QD switches from the neutral to the charged one.
The charge-dependent component $F_Q$ is a sizable contribution to the total force $F$, shown in the second panel of Fig.~\ref{Fourier} from the top. All other components of $F$ besides $F_Q$ are assumed to be conservative forces.
The dissipated energy $E_{\mathrm{diss}}$ is tied to the mechanical work consumed by the tip; it can be calculated by integrating the total force over the path given by $\Delta z(t)$.
Alternatively, because all forces except $F_Q$ are conservative, $E_{\mathrm{diss}}$ can also be calculated by integrating $F_Q dz$ alone, as indicated in the top panel of Fig.~\ref{Fourier}.

First, let us discuss the origin of the strong fluctuations observed in the dissipation signal $E_{\mathrm{diss}}$, which is especially pronounced when the measuring time equals only few oscillation periods.
The dissipation can be understood in terms of the Coulomb force $F_Q$ affecting the tip dynamics. Importantly, $F_Q$ does not change sign, being always attractive in our case.
Consequently, it accelerates the probe when the probe is approaching the surface but slows it down when it is retracting. In general, the occurrence of the  $F_Q$ is not synchronized with the motion of the probe.
The accelerating and decelerating effects cancel each other if the force $F_Q$ acts during the whole oscillation period, i.e. the case when $\nu_1 \gg \nu_2$. Similarly, there is no substantial net effect if $F_Q$ is almost zero because $\nu_2 \gg \nu_1$ or if $\nu_1, \nu_2 \ll f_0$.
In the last case, the QD may be either charged or neutral but tends to stay in the same state for the whole oscillation period.
The effect also  diminishes when the measuring time becomes sufficiently large to average out this (de)acceleration over many periods. 
However, under certain conditions the net action on the probe motion is not completely compensated and the probe becomes accelerated or damped, respectively. 
Consequently, the feedback loop has to take an appropriated action to correct the oscillation amplitude $A$.  This gives rise to sudden fluctuation of the dissipated energy $E_{\mathrm{diss}}$.
This effect is accentuated when the frequency of the charging $\nu_1$ and discharging $\nu_2$ processes are comparable to the resonant oscillation frequency $f_0$, cf.\ Eq.~(\ref{delta_Ediss_smallA}).

Secondly, the presence of positive mean dissipation signal $E_{\mathrm{diss}}$ is related to a~phase delay of the Coulomb force $F_Q$ with respect to the tip oscillations.
As we mentioned before, the tunneling rate $\nu_1$ of an~electron from the tip to the QD is given by Eq.~(\ref{tunneling}), which depends exponentially on the $z$-distance.
Therefore the charging process occurs more frequently when the tip is closer to the QD.
The actual charging tends to happen only some time after the tunneling conditions become favorable for it.
So even though the switching of $F_Q$ is random, it tends to be partially correlated with $\Delta z$, but with some delay behind it.
This lag of the charge behind tip oscillations ensures that the Coulomb force $F_Q$ resulting from the charged state acts on the tip more often when the tip goes up than when it goes down.
One effect of $F_Q$ on the probe motion, either acceleration or damping, thus prevails over the other.
This gives rise to non-conservative force component introducing non-zero dissipated energy $E_{\mathrm{diss}}$. In particular, if the Coulomb force is attractive, as we consider in our example, there is positive dissipation.
Fig.~\ref{Fourier}, top panel, shows a typical case in which there is a total energy loss (positive dissipation) over two oscillation cycles, although there is negative dissipation during the first cycle.

We should note, the dissipation signal diminishes  if the force $F_Q$ acts during the whole oscillation period, as is the case when $\nu_1 \gg \nu_2$, or if the phase shift between $F_Q$ and $z$ is negligible because $\nu_1 \gg f_0$.
On the other hand, the signal is maximal  when both tunneling rates and the oscillation frequency are comparable, $\nu_1 \approx \nu_2 \approx f_0$. Then the force $F_Q$ switches frequently during one oscillation cycle.
To justify the condition  $\nu_1 \approx \nu_2 \approx f_0$ for the maximum dissipation signal more rigorously, we derived an analytic expression for the dissipated energy $E_{\mathrm{diss}}$ (for details see Appendix B):
\begin{equation}\label{Ediss}
E_{\mathrm{diss}} = -\frac{\partial F}{\partial q} \frac{2\pi^2 A^2 \beta f_0 \bar{\nu}_1 \nu_2}{(\bar{\nu}_1 + \nu_2) \left[ (\bar{\nu}_1 + \nu_2)^2 + (2\pi f_0)^2 \right]};
\end{equation}
from the expression above, we see that the dissipated energy is proportional to the force derivative with respect to charge.  From a~detailed analysis we can also see that the dissipated energy tends to be large when $\bar{\nu}_1 \approx \nu_2 \approx f_0$, where $\bar{\nu}_1 = \nu_1(z_{\mathrm{eq}},V)$.

From the discussion above, we can deduce that the characteristic shape and observed instabilities in Kelvin parabola encode temporal information about the charge states. Thus it can be seen as a complementary tool to pump-probe STM experiments \cite{Loth_Science10}, but providing only qualitative information about characteristic tunneling rates $\nu_1, \nu_2$, (i.e.\ lifetime of generated charge state). In principle, we have established a~set of three equations (\ref{DeltaF},\ref{Ediss},\ref{FluctAndDissip}), which could be employed to determine e.g.\ the characteristic rates $\nu_1, \nu_2$. Nevertheless, this is not immediately possible, because the equations contain more unknown variables such as $\beta$ or derivatives of force. On the other hand, some of these parameters could be perhaps estimated from independent measurements on given QD system (e.g.\ $\beta$ from current measurement). However, more elaboration on the strategy is beyond the scope of this paper.

\section{Conclusions}
In conclusion, we discussed in detail the temporal response of a dynamical AFM probe to charge-state switching in QDs at different time scales.
We presented numerical simulations that captured the coupled dynamics of both the switching charge states and the oscillating probe.
We tested two complementary approaches to the simulation: a stochastic (based on pseudo-random decisions at each step) and a deterministic one (based on numerical solution of differential equations for mean values).
The analysis reveals that the presence of the resonance tunneling between the tip and QD ($\nu_1$) and between the QD and the substrate ($\nu_2$) gives rise to the instabilities in frequency shift  $\Delta f$ and the enhanced dissipated energy $E_{\mathrm{diss}}$ under certain conditions.

In addition, we derived approximate analytic formulas for the frequency shift and the dissipated energy in the limit of small amplitudes.
These formulas allows us to relate the frequency shift $\Delta f$, its fluctuation $\delta \left( \frac{\Delta f}{f} \right)$ and dissipation $E_{\mathrm{diss}}$  to the characteristic rate parameters that control the charging and discharging process, i.e.\ to the electron tunneling rates $\nu_1$ (tip--QD) and $\nu_2$ (QD--substrate). Firstly, we found that the observed frequency shift $\Delta f$ can be much larger than frequency shift corresponding to the permanently charged QD. This effect is maximized when the tunneling rates  $\nu_1$ and $\nu_2$ are of comparable magnitude. 
Secondly, the dissipated energy $E_{\mathrm{diss}}$  and the frequency-shift fluctuations $\delta \left( \frac{\Delta f}{f} \right)$ are  enhanced under the condition $\nu_1 \approx \nu_2 \approx f_0$. Thirdly, the frequency-shift fluctuation magnitude $\delta \left( \frac{\Delta f}{f} \right)$ is proportional to the square root of the expectation value of the energy dissipation $E_{\mathrm{diss}}$. Finally we discussed how the characteristic shape and observed instabilities in Kelvin parabolas encode information about temporal variations of QD charge states. 
We believe that these features can be, in principle, exploited in future research to obtain more quantitative information concerning the dynamical properties of chargeable QDs from Kelvin probe measurements. 

\section{Acknowledgement}
We thank M.~\v{S}vec, J.~Berger, and J.~Repp for valuable discussions. This work was financially supported by a Czech Science Foundation grant no.\ 14-02079S. 

\appendix

\section{Frequency shift}

We are now going to demonstrate the origin of the ``frequency shift overshoot'' and of the dissipation signal by deriving analytic formulas for both the frequency shift and energy dissipation under certain approximations.
Our goal is not finding completely general analytic formulas, which would be able to replace the numerical simulation. Instead, we want  to understand qualitatively the influence of the three time scales $f_0$, $\nu_1$ and $\nu_2$ on the measurement.
We will assume a case in which the deterministic model is a ``good enough'' description. Therefore we disregard the stochastic nature of the charging process. We will consider only the small amplitude limit of the cantilever oscillations,
so that we can restrict the changes of the short-range force and tunneling probabilities to the first order in Taylor expansion. It means that we only consider constant terms and terms linear in the position $z$. With this approximation, tip oscillation is well described by a sinusoidal function
\begin{equation}\label{dzt}
\Delta z(t) \approx A \cos (2\pi f t) = A \Re [ \exp(2\pi i f t)]
\end{equation}
as a function of time $t$ (where $\Delta z = z - z_{\mathrm{eq}}$) We have arbitrarily chosen $t=0$ in such a way that $\Delta z(t)$ is the cosine function with a zero phase shift.
In what follows, the phase of other periodically oscillating quantities will be given relative with respect to the phase of $\Delta z(t)$.
We will look for a harmonic solution to describe the temporary changes of the charge $q(t)$ too:
\begin{equation}\label{q_harm}
q(t) = \bar{q} + A_q \cos(2\pi f t + \phi_q) = \bar{q} + \Re \left [ \hat{q} \exp(2\pi i f t ) \right].
\end{equation}
In the second form of the above expression, we have introduced the complex amplitude $\hat{q} = A_q \exp (2\pi i \phi_q)$.
The complex formalism will be more convenient for the next steps of the derivation than working with sine and cosine functions.
The $z$-dependence of the tunneling rate, Eq.~(\ref{tunneling}), can be rewritten as
\begin{equation}
\nu_1 (z) = \bar{\nu}_1 \exp(-\beta \Delta z ),
\end{equation}
where $\bar{\nu}_1 = \nu_1 (z_{\mathrm{eq}} )$. 
In the small amplitude approximation, the $z$-dependence of $\nu_1$ can be linearized as
\begin{equation}\label{nu1_lin}
\nu_1 = \bar{\nu}_1 (1 - \beta \Delta z ).
\end{equation}
With $\Delta z$ given by Eq.~(\ref{dzt}), Eq.~(\ref{nu1_lin}) can be rewritten as
\begin{equation}\label{nu1}
\nu_1 = \bar{\nu}_1 \left(1 - \beta A \Re \left[ \exp(2\pi i f t) \right] \right)
\end{equation}
and if we then substitute Eq.~(\ref{q_harm}) and Eq.~(\ref{nu1}) into Eq.~(\ref{Qrate}) while
neglecting a term proportional to $A\times A_q$ (justified in the small amplitude limit), we get
\begin{equation}\label{qcomplex}
\Re \left[ 2\pi i f \hat{q} \exp(2\pi i f t ) \right] = \bar{\nu}_1 \left(1 - \beta A \Re [ \exp(2\pi i f t )] \right) (1 - \bar{q}) - \bar{\nu}_1 \Re [ \hat{q} \exp(2\pi i f t )] - \nu_2 \left(\bar{q} + \Re [ \hat{q} \exp(2\pi i f t )] \right).
\end{equation}
After rearrangement, Eq.~(\ref{qcomplex}) becomes
\begin{equation}
\Re \left[ (2\pi i f + \bar{\nu}_1 + \nu_2 ) \hat{q} \exp(2\pi i f t ) \right] = -\beta A \bar{\nu}_1 (1 - \bar{q}) \Re \left[ \exp(2\pi i f t) \right] + \bar{\nu}_1 - ( \bar{\nu}_1 + \nu_2 ) \bar{q}
\end{equation}
The last equation will be satisfied for arbitrary $t$ if
\begin{equation}\label{qbar}
\bar{q} = \frac{\bar{\nu}_1}{\bar{\nu}_1 + \nu_2}
\end{equation}
and
\begin{equation}\label{qhat}
\hat{q} = -\beta A \frac{\bar{\nu}_1 \nu_2}{(\bar{\nu}_1 + \nu_2)(\bar{\nu}_1 + \nu_2 + 2\pi i f)}.
\end{equation}
The time dependence of the charge $q(t)$ in the small amplitude approximation can be thus obtained by substituting the expressions Eq.~(\ref{qbar}) and Eq.~(\ref{qhat}) into Eq.~(\ref{q_harm}).

The frequency shift measured in AFM is given by \cite{Giessibl}
\begin{equation}
\Delta f = -\frac{f_0}{kA^2} \langle F \Delta z \rangle,
\end{equation}
assuming $\Delta f \ll f_0$ and thus $f \approx f_0$.
The angle brackets denote simultaneous temporal and ensemble averaging. Showing only the time averaging explicitly, we can write
\begin{equation}\label{DeltaFfromFt}
\Delta f = -\frac{f_0^2}{kA^2} \int_0^{T_o} F(t) \Delta z(t) dt = -\frac{f_0^2}{kA} \int_0^{T_o} F(t) \cos(2\pi f_0 t) dt,
\end{equation}
where $T_o = 1/f_0$ is the period of the cantilever oscillation.
As the interaction force $F(t)$ felt by the oscillating tip depends on the tip position $z(t)$ and on the quantum-dot charge $q(t)$.
For small amplitudes, we can linearize $F(z,q)$ and write (retaining the $f \approx f_0$ approximation from now on)
\begin{equation}
F(t) = \frac{\partial F( z = z_{\mathrm{eq}}, \bar{q})}{\partial z} \Delta z(t) + \frac{\partial F( z_{\mathrm{eq}}, q = \bar{q})}{\partial q} \Re [\hat{q} \exp( 2\pi i f_0 t )].
\end{equation}
The partial derivatives should be derived from a particular model of the interaction force, for instance from Eqs.~(\ref{LJforce}-\ref{CoulombForce}) in our case.
Let us rewrite the time dependence of $F(t)$ in terms of trigonometric functions.
\begin{equation}\label{Ft}
F(t) = \frac{\partial F}{\partial z} A  \cos(2\pi f_0 t) + \frac{\partial F}{\partial q} [\Re(\hat{q}) \cos(2\pi f_0 t) - \Im(\hat{q}) \sin(2\pi f_0 t)].
\end{equation}
Here, we have abbreviated the notation for partial derivatives by omitting the arguments in parentheses.
We can finally insert Eq.~(\ref{Ft}) into Eq.~(\ref{DeltaFfromFt}) to obtain
\begin{equation}
\Delta f = -\frac{f_0}{2k} \left(\frac{\partial F}{\partial z} + \frac{\partial F}{\partial q} \frac{\Re(\hat{q})}{A} \right).
\end{equation}
Using Eq.~(\ref{qhat}),
\begin{equation}\label{DeltaF_final}
\Delta f = \frac{f_0}{2k} \left( -\frac{\partial F}{\partial z} + \frac{\partial F}{\partial q} \frac{\beta \bar{\nu}_1 \nu_2}{(\bar{\nu}_1 + \nu_2)^2 + (2\pi f_0)^2} \right).
\end{equation}
The first term in the round brackets on the right-hand side of the above equation corresponds to frequency shift that would be observed for a stationary charge $q = \bar{q}$.
The second term is the part of the frequency shift contributed by the charge dynamics on the QD. This second term may lead to the frequency shift ``overshoot'' observed in the results of our simulations
when it becomes large at $\nu_1 \approx \nu_2 \gg f_0$.

\section{Dissipation}

Energy dissipation per cycle can be expressed as
\begin{equation}\label{W}
E_{\mathrm{diss}} = -W = -\int_0^{T_o} F(t) \frac{dz}{dt} dt = 2\pi A f_0 \int_0^{T_o} F(t) \sin(2\pi f_0 t) dt.
\end{equation}
The dissipation, apart from a constant contribution from the finite quality factor, originates only from the electrostatic force related to the switching charge, because all other forces are conservative.
Consequently, in the small amplitude approximation
\begin{equation}\label{W_from_q}
E_{\mathrm{diss}} = 2\pi A f_0 \frac{\partial F}{\partial q} \int_0^{T_o} \Re[\hat{q} \exp(2\pi i f_0 t)] \sin(2\pi f_0 t) dt = - \pi A \frac{\partial F}{\partial q} \Im(\hat{q}).
\end{equation}
So finally, the dissipation will be
\begin{equation}\label{Ediss_final}
E_{\mathrm{diss}} = -\frac{\partial F}{\partial q} \frac{2\pi^2 A^2 \beta f_0 \bar{\nu}_1 \nu_2}{(\bar{\nu}_1 + \nu_2) \left[ (\bar{\nu}_1 + \nu_2)^2 + (2\pi f_0)^2 \right]}.
\end{equation}
The above expression for dissipated energy assumes its maximal absolute value when
\begin{equation}\label{maxcondition}
\bar{\nu}_1 = \nu_2 = \pi f_0.
\end{equation}
To proof this statement, consider that according to Eq.~(\ref{Ediss_final}), the dissipated energy is proportional to
\begin{equation}\label{Ediss_proportionality}
E_{\mathrm{diss}} \propto \frac{f_0 \bar{\nu}_1 \nu_2}{(\bar{\nu}_1 + \nu_2) \left[ (\bar{\nu}_1 + \nu_2)^2 + (2\pi f_0)^2 \right]}
= \frac{f_0 \left[(\bar{\nu}_1 + \nu_2)^2 - (\bar{\nu}_1 - \nu_2)^2\right]}{4 (\bar{\nu}_1 + \nu_2) \left[ (\bar{\nu}_1 + \nu_2)^2 + (2\pi f_0)^2 \right]}.
\end{equation}
The second term in the square bracket in the numerator of the rightmost side of Eq.~(\ref{Ediss_final}) always reduces the absolute value of the whole expression for $E_{\mathrm{diss}}$.
This term can never be negative, it must be at least zero, in which case the whole expression will be maximized.
To make the term $(\bar{\nu}_1 - \nu_2)^2$ zero, one should require
\begin{equation}
\bar{\nu}_1 = \nu_2 \equiv \nu.
\end{equation}
Now, Eq.~(\ref{Ediss_proportionality}) simplifies to
\begin{equation}\label{Ediss_when_nu_equal}
E_{\mathrm{diss}} \propto \frac{f_0 \nu^2}{8\nu \left[\nu^2 + (\pi f_0)^2 \right]} = \frac{f_0 \nu}{8\left[(\nu - \pi f_0)^2 + 2\nu \pi f_0 \right]} \\ = \frac{1}{\frac{8(\nu - \pi f_0)^2}{f_0 \nu} + 16\pi}.
\end{equation}
The last expression obviously maximizes if the first term in the denominator becomes zero, which happens when $\nu = \pi f_0$. This completes the proof of Eq.~(\ref{maxcondition}).

\section{Fluctuations}

Let us now turn our attention to the random fluctuations seen in the measurable quantities $\Delta f$ and $E_{\mathrm{diss}}$.
These fluctuations become apparent in results obtained within the stochastic model, see Figs.~\ref{dfResults} and \ref{StatisticsFig}.
We will now try to estimate the dependence of these fluctuations on $f_0 = 1/T_o$, $\nu_1$, and $\nu_2$ quantitatively.
Besides these parameters, the fluctuations obviously depend on the number of periods $N$ (i.e.\ on the time $t_m = NT_o = N/f_0$) over which the results for $\Delta f$ and $E_{\mathrm{diss}}$ are
 taken by the measurement of a single point of the $\Delta f(V)$ curve.
We will characterize the fluctuations quantitatively by root mean square deviations:
\begin{equation}\label{delta_df}
\delta(\Delta f) = \sqrt{\left\langle \left( \Delta f - \left\langle \Delta f \right\rangle \right)^2 \right\rangle},
\end{equation}
\begin{equation}\label{delta_Ediss}
\delta E_{\mathrm{diss}} = \sqrt{\left\langle \left( E_{\mathrm{diss}} - \left\langle E_{\mathrm{diss}} \right\rangle \right)^2 \right\rangle },
\end{equation}
where
\begin{equation}\label{df_stoch}
\Delta f = -\frac{f_0^2}{NkA^2} \int_0^{NT_o} F(t) z(t) dt = -\frac{f_0^2}{Nk} \int_0^{NT_o} F(t) \cos(2\pi f_0 t) dt,
\end{equation}
\begin{equation}\label{Ediss_stoch}
E_{\mathrm{diss}} = -\frac{1}{N} \int_0^{NT_o} F(t) \frac{dz}{dt} dt = \frac{2\pi A f_0}{N} \int_0^{NT_o} F(t) \sin(2\pi f_0 t) dt,
\end{equation}
cf.~Eq.~(\ref{DeltaFfromFt}) and Eq.~(\ref{W}).
As before, the angle brackets denote ensemble averaging.

We will start with general considerations about $\delta(\Delta f)$ and $\delta E_{\mathrm{diss}}$, valid for arbitrary amplitudes,
and we will then switch to the small amplitude approximation to derive analytic formulae for $\delta(\Delta f)$ and $\delta E_{\mathrm{diss}}$

The non-random conservative components of the total force $F$ do not contribute to either $\delta(\Delta f)$ or $\delta E_{\mathrm{diss}}$.
Furthermore, we now work within the stochastic model.  Thus we assume that the charge can only be either $q=0$ or $q=1$.
Therefore the charge dependence of the force can be simplified as
\begin{equation}
F_Q(z,q) = F_Q(z)q,
\end{equation}
where $F_Q(z)$ is the charge-dependent component for $q=1$.
As we will shortly see, the restriction of $q$ to only two possible values will be a great help in the evaluation of the fluctuations.
Furthermore, we neglect the $z$ dependence of $F_Q(z)$ over the range of the oscillating tip.\footnote{
This simplification need not be valid for large amplitudes. We could postpone introducing it until we fully embrace the small amplitude approximation.
Working with $z$-dependent $F_Q$ would prevent us from extracting $F_Q$ in front of the integrals in the subsequent steps.
Instead, we would have to keep $F_Q(z) = F_Q\left(z_{\mathrm{eq}}+A\cos(2\pi f_0 t)\right)$ inside the integrals. Otherwise, the derivation would not change.
At the point when we eventually start using the small-amplitude approximation in a consistent way, we would set $F_Q(z)=F_Q$ anyway.}
It allows us to rewrite Eqs.~(\ref{df_stoch}) and (\ref{Ediss_stoch}) in terms of $q(t)$ as
\begin{equation}\label{df_q}
\Delta f = -\frac{F_Q f_0^2}{NkA} \int_0^{NT_o} q(t) \cos(2\pi f_0 t) dt,
\end{equation}
\begin{equation}\label{Ediss_q}
E_{\mathrm{diss}} = \frac{2\pi F_Q A f_0}{N} \int_0^{NT_o} q(t) \sin(2\pi f_0 t) dt.
\end{equation}
We now substitute Eq.~(\ref{df_q}) into Eq.~(\ref{delta_df}) and Eq.~(\ref{Ediss_q}) into Eq.~(\ref{delta_Ediss}) to get
\begin{equation}\label{delta_df_q}
\delta (\Delta f) = \frac{f_0^2 F_Q}{NkA} \sqrt{\int_0^{NT_o} dt \int_0^{NT_o} dt' C_q(t,t') \cos(2\pi f_0 t) \cos(2\pi f_0 t')}
\end{equation}
and
\begin{equation}\label{delta_Ediss_q}
\delta E_{\mathrm{diss}} = \frac{2\pi f_0 F_Q A}{N} \sqrt{\int_0^{NT_o} dt \int_0^{NT_o} dt' C_q(t,t') \sin(2\pi f_0 t) \sin(2\pi f_0 t')} ,
\end{equation}
where
\begin{equation}\label{Ctt}
C_q(t,t') = \bigl\langle \bigl( q(t) - \left\langle q(t) \right\rangle \bigr) \bigl( q(t') - \left\langle q(t') \right\rangle \bigr) \bigr\rangle = \langle q(t) q(t') \rangle - \langle q(t) \rangle \langle q(t') \rangle
\end{equation}
is the correlation function of charge deviation from the mean charge value.
Because the charge may only switch between 0 and 1, we can interpret $\langle q(t)\rangle$ as the probability that $q(t) = 1$. Moreover, $\langle q(t) q(t') \rangle$ equals the probability that $q=1$ at both times $t$ and $t'$.
For $t' \geq t$, it can be expressed as
\begin{equation}\label{q_condit}
\langle q(t) q(t') \rangle = \langle q(t) \rangle \bigl\langle q(t') \big| \: q(t)=1 \bigr\rangle ,
\end{equation}
where $\langle q(t') | \, q(t) = 1 \rangle$ is the conditional probability that $q(t') = 1$ assuming $q(t) = 1$.
This conditional probability can be found by solving a differential equation of the type of Eq.~(\ref{Qrate}) for $q(t')$ in time $t' > t$, with the initial condition at $t$ defined as $q(t)=1$.
Let us denote such conditional probability or conditional mean charge value at $t'$ by $q_t(t')$, in order to distinguish it from both the unconditioned mean value $\langle q(t') \rangle$ and from a particular (non-averaged) instance of the random variable $q(t')$.
Briefly, we denote $q_t(t') \equiv \langle q(t') | \, q(t) = 1 \rangle$.
With such a careful notation, Eq.~(\ref{Qrate}) can be rewritten as
\begin{equation}\label{diffeq_qt}
\left. \frac{q_t(t')}{dt'} = \nu_1 (t') - \bigl( \nu_1(t') + \nu_2(t') \bigr) q_t(t') \: \right| \: q_t(t)=1 ; \: t' \geq t ,
\end{equation}
stating now every possible time dependence explicitly.
With the solution $q_t(t')$ of the above equation available, we can easily evaluate the correlation function $C_q$ by combining Eq.~(\ref{Ctt}) and Eq.~(\ref{q_condit})
if we also know $\langle q(t) \rangle$:
\begin{equation}
 C_q(t,t') = \langle q(t) \rangle \bigl( q_t(t') - \langle q(t') \rangle \bigr) \: \mathrm{for} \: t' \geq t.
\end{equation}
Actually, we can use the solution of the differential equation for $q_t(t')$ to find $\langle q(t) \rangle$ too. 
The charge dynamics is driven by the periodic oscillations of the cantilever, so we can expect that after long enough time, the mean value of $q_t(t')$ will converge to a stationary periodic solution:
\begin{equation}
\lim_{N\rightarrow\infty} q_t(NT_o+t') = \langle q(t') \rangle.
\end{equation}
Considering the above explained prescription to derive the correlation function, it would be convenient if we needed to evaluate $C_q(t,t')$ only for $t' \geq t$.
This can be indeed achieved thanks to the fact that the integrands in Eq.~(\ref{delta_df_q}) and Eq.~(\ref{delta_Ediss_q}) are symmetric with respect to the exchange $t \leftrightarrow t'$.
\begin{equation}\label{delta_df_corr}
\delta (\Delta f) = \frac{f_0^2 F_Q}{NkA} \sqrt{2 \int_0^{NT_o} dt \int_t^{NT_o} dt' C_q(t,t') \cos(2\pi f_0 t) \cos(2\pi f_0 t')}
\end{equation}
and
\begin{equation}\label{delta_Ediss_corr}
\delta E_{\mathrm{diss}} = \frac{2\pi f_0 F_Q A}{N} \sqrt{2 \int_0^{NT_o} dt \int_t^{NT_o} dt' C_q(t,t') \sin(2\pi f_0 t) \sin(2\pi f_0 t')} .
\end{equation}
What remains to be done so as to find the fluctuations defined by Eq.~(\ref{delta_df}) and Eq.~(\ref{delta_Ediss}) is to solve Eq.~(\ref{diffeq_qt}) for $q_t(t')$, thus finding the correlation function $C_q(t,t')$,
substitute $C_q(t,t')$ into Eq.~(\ref{delta_df_corr}) and Eq.~(\ref{delta_Ediss_corr}) and evaluate the integrals.
In order to solve Eq.~(\ref{diffeq_qt}), however, we have to specialize to a particular model for $\nu_1(t)$ and $\nu_2(t)$.
At this point, the following observation may be worth emphasizing: Although the existence of fluctuations in measurable quantities like $\Delta f$ or $E_{\mathrm{diss}}$ is a consequence of the \emph{stochastic} nature of the charging and discharging process,
a crucial step needed to quantify these fluctuations consists in solving the differential equation Eq.~(\ref{diffeq_qt}), which can be done numerically within what we call the \emph{deterministic} approach.
But we would also like to find an analytical expression for the fluctuations, and in order to do so, we will
go back to the small amplitude approximation. We again assume a time independent discharging rate $\nu_2$.
Moreover, we are going to content ourselves with the lowest-order term from the expansion of $\delta(\Delta f)$ and $\delta E_{\mathrm{diss}}$ in powers of the amplitude.
For that, we will need only the amplitude-independent part of $C_q(t,t')$, which can be obtained by assuming that the charging rate $\nu_1$, formerly given by Eq.~(\ref{nu1}), becomes time-independent too, $\nu_1 = \bar{\nu}_1$.
We now have the solution of Eq.~(\ref{diffeq_qt}) (with time-independent rate constants)
\begin{equation}
q_t(t') = (1 - \bar{q}) e^{ -(\bar{\nu}_1 + \nu_2) (t' - t) } +  \bar{q}. 
\end{equation}
For $C_q(t,t')$, it follows that
\begin{equation}
C_q(t,t') = \bar{q} (1 - \bar{q}) e^{ -(\bar{\nu}_1 + \nu_2) (t' - t) } = \frac{\bar{\nu}_1 \nu_2}{(\bar{\nu}_1 + \nu_2)^2} \exp [ -(\bar{\nu}_1 + \nu_2) (t' - t) ].
\end{equation}
When evaluating the double integrals in Eq.~(\ref{delta_df_corr}) and Eq.~(\ref{delta_Ediss_corr}), we will consider only terms proportional to the measurement time $NT_o$.
Because the square root is taken of the double integrals and there is a prefactor that involves $1/N$ in front of the square root,
such terms will eventually decrease as $1/\sqrt{N}$ with increasing $N$. All other possible terms would decrease faster and they can be thus neglected unless $N$ is too small.
We should also note that the terms which decrease with $N$ faster than $1/\sqrt{N}$ are sensitive to the exact initial and final conditions of the measurement,
in particular on the phase of $\Delta z(t)$ at $t=0$ (the beginning of the measurement) and at $t=NT_o$ (the end of the measurement).
While we have chosen $\Delta z(t) = A\cos(2\pi f_0 t)$, so $\Delta z (0)= +A$, and we have furthermore assumed $N$ to be a natural number, so $\Delta z(NT_o) = +A$ too,
this particular choice is obviously arbitrary and it cannot be usually controlled in a realistic experimental setup anyway.
Neglecting terms which are sensitive to this choice is therefore justified. With this last approximation, final expressions for the fluctuations can be given.
\begin{equation}\label{delta_df_smallA}
\delta (\Delta f) \approx \frac{f_0^2 F_Q}{NkA} \sqrt{\frac{\bar{\nu}_1 \nu_2 NT_o}{ (\bar{\nu}_1 + \nu_2) \left[ (\bar{\nu}_1 + \nu_2)^2 + (2\pi f_0)^2 \right]}}
= \frac{\sqrt{f_0^3} F_Q}{\sqrt{N}kA} \sqrt{\frac{\bar{\nu}_1 \nu_2}{ (\bar{\nu}_1 + \nu_2) \left[ (\bar{\nu}_1 + \nu_2)^2 + (2\pi f_0)^2 \right]}}
\end{equation}
and
\begin{equation}\label{delta_Ediss_smallA}
\delta E_{\mathrm{diss}} \approx \frac{2\pi f_0 F_Q A}{N} \sqrt{\frac{2\pi f_0 \bar{\nu}_1 \nu_2 NT_o}{ \left(\bar{\nu}_1 + \nu_2\right)^2 \left[ (\bar{\nu}_1 + \nu_2)^2 + (2\pi f_0)^2 \right]}}
= \frac{\sqrt{(2\pi)^3} F_Q A}{\sqrt{N}} \sqrt{\frac{f_0 \bar{\nu}_1 \nu_2 }{ (\bar{\nu}_1 + \nu_2) \left[ (\bar{\nu}_1 + \nu_2)^2 + (2\pi f_0)^2 \right]}}.
\end{equation}
Note that the magnitude of relative fluctuations of the frequency shift scales with the rate and frequency constants $\bar{\nu}_1$, $\nu_2$, and $f_0$ as the square root of the dissipation,
as follows from the combination of Eq.~(\ref{delta_df_smallA}) and Eq.~(\ref{Ediss_final}) (noting that $\partial F / \partial q = F_Q$):
\begin{equation}\label{relative_delta_df}
\delta \left( \frac{\Delta f}{f} \right) = \frac{\sqrt{\partial F / \partial q}}{\sqrt{2N}\pi kA^2 \sqrt{\beta}} \sqrt{\left|E_{\mathrm{diss}}\right|}.
\end{equation}



\bibliography{refs}

\end{document}